\documentclass[pre,aps,showpacs]{revtex4}

\usepackage{amsmath}
\usepackage{graphicx}
\usepackage{amssymb}

\newcommand{\vt}{v_{th}}
\newcommand{\bv}{{\bf v}}

\begin{document}

\title{Fluctuating hydrodynamics in a vertically vibrated granular fluid with gravity}
\author{Giulio Costantini}
\affiliation{CNR-ISC and Dipartimento di Fisica, Universit\`a Sapienza - p.le A. Moro 2, 00185, Roma, Italy}
\author{Andrea Puglisi}
\affiliation{CNR-ISC and Dipartimento di Fisica, Universit\`a Sapienza - p.le A. Moro 2, 00185, Roma, Italy}

\begin{abstract}
We investigate hydrodynamic fluctuations in a $2D$ granular fluid
excited by a vibrating base and in the presence of gravity, focusing
on the transverse velocity modes. Since the system is inhomogeneous,
we measure fluctuations in horizontal layers whose width is smaller
than the characteristic hydrodynamic lengths: they can be considered
as almost-homogeneous subsystems. The large time decay of
autocorrelations of modes is exponential and compatible with vorticity
diffusion due to shear viscosity, as in equilibrium fluids. The
velocity structure factor, which strongly deviates from the
equilibrium constant behavior, is well reproduced by an effective
fluctuating hydrodynamics described by two noise terms: 
the first associated with vorticity diffusion and the second with the local energy exchange,
which have internal and external character, respectively.
\end{abstract}


\maketitle

\section{Introduction}
Granular systems are ubiquitous on Earth: understanding those
materials is essential for the optimization of a variety of industrial
processes in the pharmaceutical, food, cosmetic, chemical, petroleum,
polymer and ceramic industries~\cite{intro1,JNB96b}, it is also 
propaedeutic for modeling geophysical phenomena such as sediment
fluidization in sudden
landslides~\cite{intro_geo1,intro_geo2,intro_geo3} and soil
mechanics~\cite{intro_soil}. Finally, it provides to be a challenge
for the statistical physics of out-of-equilibrium complex
materials~\cite{K99,kudrolli,irrevers,biroli,lengthscale}. Fluctuations
in granular media are usually non-negligible because of their
inherently ``small'' nature: they are usually composed of a few
thousands particles, sometimes even less. On the other side
equilibrium statistical mechanics, which is equipped with tools to
describe large scale fluctuations of finite
systems~\cite{LandauFisStat}, is not adequate for systems with
non-conservative interactions, such as granular fluids undergoing
inelastic collisions. A deep understanding of the single-particle
velocity fluctuations has been obtained in recent years, through the
study of the granular Boltzmann equation and the introduction of
several effective models~\cite{esipov,NE98,WM96,PLMPV98}. Fluctuations
involving more than one degree of freedom, i.e. correlations, are
instead less understood: the reason is that they not always have an
equilibrium counterpart. For instance, velocity correlations are ruled
out in homogeneous steady elastic fluids, while they play an important
role in homogeneous steady inelastic
gases~\cite{NEBO97,NETP99,TPNE01,BMP02b}. Important progresses on the
study of velocity correlations have been achieved for hydrodynamic
(i.e. ``slow'') modes, in situations where a good separation of length
and time scales is observed. While phenomenological fluctuating
hydrodynamics has been used in the first study of velocity
correlations in homogeneous granular gases~\cite{NEBO97}, a more
rigorous treatment through projection formalism from microscopic
models has been recently obtained~\cite{BMG09}. One of the main
results, seen both in the theory and in simulations~\cite{BGM08,CP10},
is that the hydrodynamic noise is not white and does not satisfy the
second kind Fluctuation-Dissipation Relation (FDR). These two
deviations from the equilibrium fluctuating hydrodynamics are
observable but, however, relatively small in homogeneous systems. On
the other side, a certain sensitivity to the particular modelling of
energy injection has been observed~\cite{VPBWT06,MGT09}. It is
therefore tempting to extend this study to more realistic energy
injection models, with the objective of providing a framework for
future experiments and more formal analytical treatment. Here, in
particular, we are interested in simulating a typical experimental
setup, that is a 2D vertical box, under gravity, whose vibrating base
acts as a thermal reservoir~\cite{BRM01,BBPV01,VPBWT06,PCMP03}. The loss of
kinetic energy, due to inelastic collisions in the bulk, is
counterbalanced by the random floor vibration which injects energy to
particles bouncing on it. This setup, which has received careful
hydrodynamic treatment~\cite{BRM01}, is characterized by non-constant
density and temperature profiles and by a steady (inhomogeneous)
heat flux~\cite{BR04}, carrying energy from the bottom region up to
the top one. Our analysis, which is purely numerical, focuses on the
velocity fields, which is zero on average, provided that convection
rolls are not present in the system~\cite{RRC00,PBMP04}. Fluctuations
of the shear modes (vertical velocity modulated in the horizontal
direction at wavelength $k$) are measured in subsystems small enough
to be considered ``homogeneous'': a subsystem here is a horizontal
layer taking the full width of the system and a fraction of its
height, smaller than the smaller characteristic length of hydrodynamic
profiles. Three main characterizations of these fluctuations are in
order: a) the large time decay of their autocorrelation $\langle
U_\perp(k,t)U_\perp(-k,0)\rangle \sim \exp[-q(k)t]$; b) the dependence
of $\langle U_\perp(k,0)U_\perp(-k,0)\rangle$ with $k$, which is a
constant for equilibrium fluids, and it is not in non-equilibrium
ones; c) the relation between the two previous quantities, which at
equilibrium is established in a simple form by the FDR. Previously,
fluctuations have been studied in this particular ``setup'' focusing
on other quantities. In~\cite{BBPV01} single particle velocity
fluctuations and density correlations at different height have been
studied. In~\cite{VPBWT06} the fluctuations of the total kinetic
energy have been investigated, showing that strong non-Gaussian
behavior is not due to strong correlations but can be simply accounted by considering inhomogeneous temperature and local Gaussian
velocity distributions. This consideration led us to the choice of
studying sub-regions of the whole system where homogeneity can be
assumed and velocity distributions are not far from the Gaussian. The
drawback of this choice is, of course, that each sub-region is an {\em
  open} system and this fact must be taken into account. In Section II
we define the model, the simulation scheme and the quantities under
study. In Section III the separation of the inhomogeneous
system into nearly homogeneous sub-systems is detailed. In Section IV we discuss
the numerical results for the points a, b and c described above. In
Section V we draw some conclusion and discuss perspectives.

\section{Model and quantities under study}
We perform Molecular Dynamics simulations of inelastic hard disks of
diameter $\sigma$ and mass $1$, enclosed in a $2D$ rectangular volume
of width $L_x$ and height $L_y$, under the presence of gravity
(pointing in the $-\hat{y}$ direction) with acceleration $g$.  The
energy is injected to the system by a thermal base: this means that a
particle colliding with the wall at $y=0$ is bounced back with its $x$
velocity component $v_x$ extracted by a Gaussian distribution with
temperature $T_b$ and with its $y$ velocity component $v_y$ extracted
with the distribution~\cite{TTKB98}
\begin{equation}
P(v_y)=\frac{v_y}{T_b} e^{-\frac{v_y^2}{2T_b}}.
\end{equation}
The ceiling of the box is elastically reflecting. Periodic boundary
conditions in the $\hat{x}$ direction are adopted. The disks collide
inelastically with usual rule dictating the instantaneous velocity
change $\bv_i \to \bv_i'$ of particle $i$ after a collision with
particle $j$:
\begin{equation}
\bv_i'=\bv_i-\frac{1+\alpha}{2}[(\bv_i-\bv_j)\cdot\hat{n}]\hat{n},
\end{equation}
where $\alpha \in [0,1]$ is the restitution coefficient ($=1$ for
elastic systems) and $\hat{n}$ is the unit vector joining the
colliding disks.  We are interested in measuring the velocity shear
mode, modulated along the $\hat{x}$ direction, which is the only one
where periodic boundary conditions are enforced. This fluctuating
variable with complex values is defined as
\begin{equation}
 U_{\perp}(k,t)= \sum_{j=1}^N v_{y,j}(t) e^{-i k x_j(t)}.
\label{Uk}
\end{equation}
where $k$ is the wave number of chosen mode, $N$ the total number of particles and $x_j(t)$ is the
$x$-coordinate of particle $j$ at time $t$. \\ The main
function under investigation is the rescaled autocorrelation of mode
$k$:
\begin{equation}
C_{\perp}(k,t)= \frac{\langle U_{\perp}(k,0)U^*_{\perp}(k,t)\rangle} {\vt^2},
\end{equation}
measured in the steady state, where $\vt^2=2T_{g,y}$ and
$T_{g,\beta}=\langle v_\beta^2\rangle$ is the ``granular temperature''
in the $\beta$ direction, with $\beta$ being $x$ or $y$. Note that
$T_{g,y}/T_{g,x}\approx 1$ in the region of interest (which includes
most of the system, excluding the boundary layers near the bottom and
top walls). We will also use the shorthand notation $U_\perp(t)\equiv
U_\perp(k_{min},t)$ and $C_\perp(t)\equiv C_\perp(k_{min},t)$ for the
largest mode $k_{min}=2\pi/L_x$. In particular we will inspect two
main features of the above correlation function, i.e.:
\begin{itemize}
\item its large time decay, focusing on the 
dependence on the parameters such as the granular temperature $T_{g,y}$ and the particle density $n$ and
on the wave number $k$;
\item the variance of the fluctuations of the modes as a function of
  $k$, $C_\perp(k,0)$, which is nothing else than the rescaled
  velocity structure factor; that quantity gives information about the
  spatial extent of velocity correlations.
\end{itemize}
Since the fluid receives energy from a boundary and gravity is
present, all hydrodynamic profiles are inhomogeneous. In order to
avoid the mixing of information coming from regions with different
densities and temperatures, in the following section we discuss how to
define sub-systems which can be assumed to be nearly homogeneous for
our purposes. Before doing that, however, we discuss what is expected
in previously studied homogeneous systems.

\subsubsection{Behavior of a homogeneous system}

For comparison, we briefly review the behavior of the above quantities
in an equilibrium homogeneous 2D system and in a granular system both
in the homogeneous cooling regime and in the homogeneous stationary
state due to uniform random driving.

At equilibrium (at temperature $T_g=v_{th}^2/2$), the Landau-Lifshitz
fluctuating hydrodynamics, based on Einstein fluctuation formula,
predicts a Langevin equation for the hydrodynamic shear modes of the
kind
\begin{equation}
\partial_t U_{\perp}(k,t)=-\nu k^2 U_{\perp}(k,t)+\xi(k,t), 
\label{LangevinLL}
\end{equation}
where $\nu$ is the kinematic viscosity and $\xi(t)$ is a white noise with zero average and 
\begin{equation}
\langle \xi(k,t) \xi(k',t')\rangle = \delta_{k',-k} \delta(t-t') 2T_gN\nu k^2.
\end{equation}
Based on this equation, one has
\begin{equation}
C_{\perp}(k,t) = \frac{N}{2} e^{-\nu k^2 t}.
\label{corrLL}
\end{equation}

In the inelastic homogeneous cooling regime, which is also dilute, 
the amplitude of
fluctuations is decaying: indeed, because of cooling, the kinematic
viscosity $\nu$ and $T_g$ decrease with time. In these systems $\nu$
can be written, in the first Sonine approximation, as
\begin{equation}
\nu=g_2(\sigma)\frac{\lambda v_{th}}{\sqrt{2\pi}} \frac{4}{1+\alpha}\Big[5-\alpha - \frac{a_2(\alpha)}{32}(19-15\alpha)\Big]^{-1}
\label{eqnu}
\end{equation}
where $g_2(\sigma)$ is the equilibrium value of the pair-correlation function, $\lambda=(n\sigma)^{-1}$ is proportional to the mean free path
and the coefficient $a_2(\alpha)$ is a function of the restitution
coefficient only (see \cite{BMG09} for details). \\ For that reason
one uses a new time-scale $\tau$ which is proportional to the
cumulative number of collisions~\cite{BRM04}: under this new
time-scale the system reaches a stationary regime. Rigorous treatment
from a fluctuating Boltzmann equation in this regime
yields~\cite{BMG09}
\begin{equation}
\partial_\tau U_{\perp}(k,\tau)=-\left(\nu k^2 - \frac{\zeta_H}{2}\right) U_{\perp}(k,\tau)+\xi'(k,\tau), 
\label{LangevinHcs}
\end{equation}
where
\begin{equation}
\zeta_H=\frac{v_{th}}{\lambda}\sqrt{\frac{\pi}{2}}(1-\alpha^2)\Big[1+ \frac{3}{16}a_2(\alpha)\Big]
\label{eqzeta}
\end{equation}
is the cooling rate (see \cite{BRM04} for a definition). The noise results to be non-white, with a correlation
\begin{equation}
\langle \xi'(k,\tau)\xi'(k',\tau') \rangle = \delta_{k',-k} 2NT_gk^2 G(|\tau-\tau'|),
\end{equation}
with $G(|s|) \neq \delta(s)$ a function which is given in details
in~\cite{BMG09}. The steady autocorrelation $C_{\perp}(\tau)$ obtained from the
above Langevin equation is not a simple exponential; anyway it has an
exponential tail at large times. In particular one obtains
\begin{align} \label{corrHcs}
C_\perp(k,0)&=\frac{N}{2} \frac{\nu_1 k^2}{\nu k^2-\zeta_H/2}\\
C_\perp(k,\tau)&=\frac{N}{2} \frac{(\nu_1+\nu_2) k^2}{\nu k^2-\zeta_H/2}e^{-(\nu k^2-\zeta_H/2)\tau} \;\;\;\;\; \tau \to \infty \label{corrHcs2}
\end{align}
where the two new coefficients $\nu_1$ and $\nu_2$ (the latter is
usually smaller than the first) are computed in~\cite{BMG09}. In the
elastic limit $\zeta_H \to 0$, $\nu_1 \to \nu$ and $\nu_2 \to 0$ and
result~\eqref{corrLL} is recovered. In the inelastic case
($\zeta_H>0$) one immediately sees that the theory is limited to $k$
large enough to have $\nu k^2-\zeta_H/2>0$: modes with smaller
wavenumbers are unstable. That condition obliges to consider systems
smaller than a critical size to avoid the instability.

Finally, we consider the case of a homogeneously driven granular gas,
as described in~\cite{NETP99}, where the hydrodynamics of a gas of
inelastic grains which receive energy by random uncorrelated velocity
kicks is studied.  In that work hydrodynamic fluctuations are
described using an effective noise that is the sum of an internal and
an external noise. The former is originated from the rapid
fluctuations of microscopic degrees of freedom and its strength can be
obtained from an FDR with respect to internal relaxation. The latter,
instead, is due to the random accelerations received by particles from
the external driving. The strength of this noise is such that, in
the steady state, it balances the energy loss due to the
collisions. The result for the shear mode with small inelasticity is a Langevin equation:
\begin{equation}
\partial_\tau U_{\perp}(k,t)=-\nu k^2 U_{\perp}(k,t)+\xi''(k,t), 
\label{LangevinTrizac}
\end{equation}
with
\begin{equation}
\langle \xi''(k,t) \xi''(k',t')\rangle=\delta_{k',-k}\delta(t-t') N T_g \left( 2\nu k^2 +\zeta_H\right).
\end{equation}
The corresponding autocorrelation $C_{\perp}(k,t)$ is written as
\begin{equation}
C_{\perp}(k,t) = \frac{N}{2} \frac{\zeta_H/2+\nu k^2}{\nu k^2} e^{-\nu k^2 t}.
\label{corrLL2}
\end{equation}
We stress that other models exist to get a spatially homogeneous and
stationary granular gas, for instance an important modification of the
last one is that introduced in~\cite{PLMPV98}, where all particles
feel the presence of a viscous bath: the fluctuating hydrodynamics for
such a model has been recently studied~\cite{GSVP11} and demonstrated
to fairly describe the experimental behavior of a quasi-2D system on a
horizontal vibrating plate. \\ An important property common to
  all the systems discussed above is the presence of spatial and
  thermal homogeneity that is, instead, absent in our system. We
  mention that fluctuating hydrodynamics in molecular (elastic) fluids
  where a non-equilibrium stationary state is obtained by imposing
  (temperature or density) gradients has been studied in the
  literature, see for instance~\cite{SC85,OS06}. In particular,
  in~\cite{SC85} the authors analyze the fluctuations of a fluid under
  a stationary heat flux in the presence of a gravity field. They show
  that, for small temperature gradients, such system can be studied
  within a Landau-Lifshitz approach postulating that FDR still holds
  locally. This implies that the transverse velocity modes are well
  described by Eq. (\ref{LangevinLL}) where all parameters and
  coefficients take their local (position-dependent) value, and
  therefore do not display any correlation, yielding a flat structure factor.

In the following we will see that the hydrodynamic fluctuations of the
granular setup considered here, which is a nearly homogeneous {\em
  sub-system} belonging to a inhomogeneous one, are, for some aspects,
well reproduced by a model similar to Eq.~\eqref{LangevinTrizac}.


\section{Quasi-homogeneous sub-systems}
\begin{figure}[htbp]
\begin{center}
\includegraphics[angle=0,width=5.5cm,clip=true]{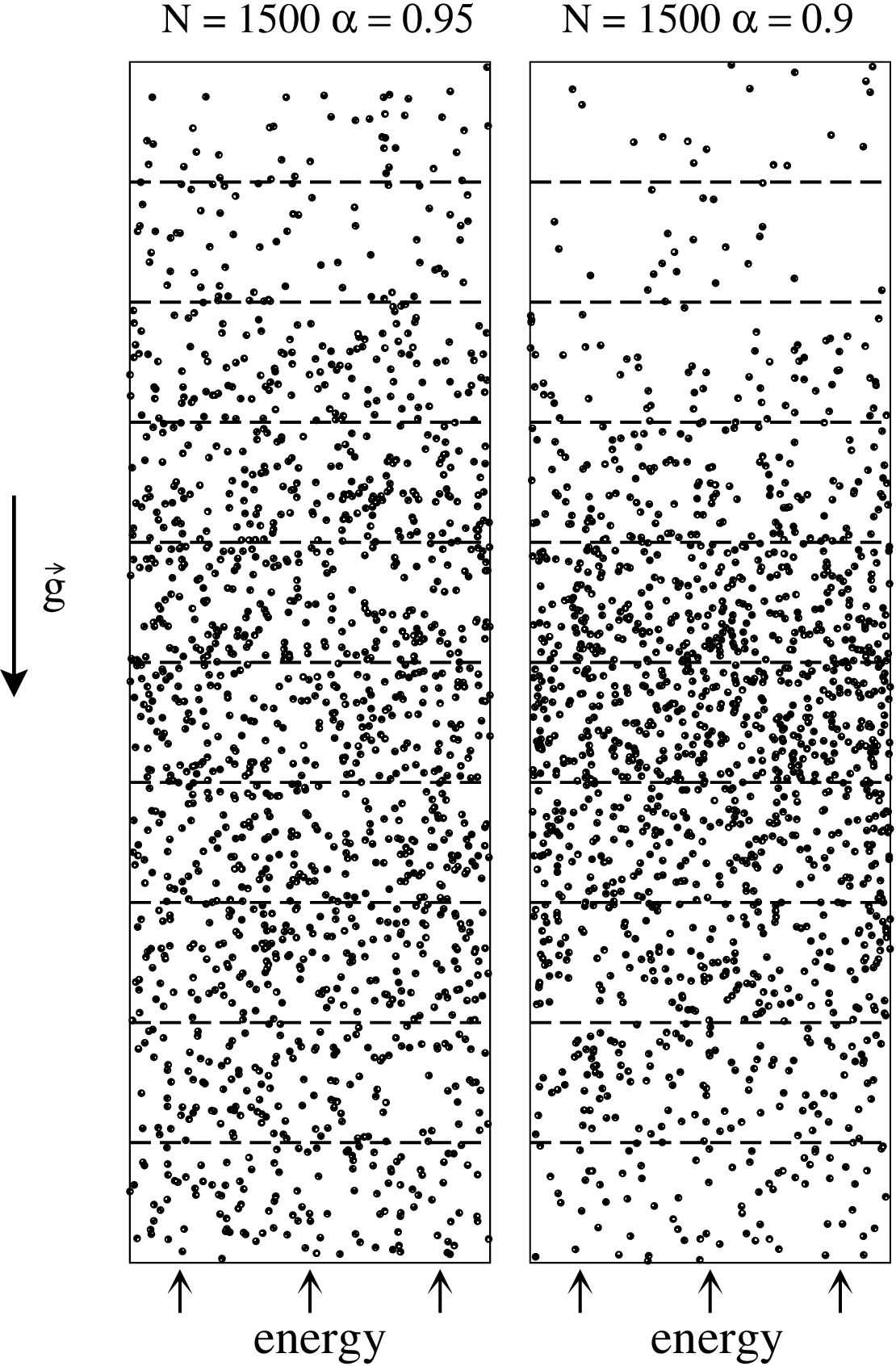}
\caption{Snapshot of the system with $N=1500$ particles for two
  different restitution coefficient $\alpha$: left $\alpha=0.95$,
  right $\alpha=0.9$. The other parameters are $L_y/\sigma=600$,
  $L_x/\sigma=180$ and $T_b/(gL_y)=2.55$ for the left panel, while
  $T_b/(gL_y)=5.1$ for the right panel.
\label{fig1}}
\end{center}
\end{figure}

The main external parameters influencing the regime are, besides the
dimensions of the box, the number $N$ of particles, the floor
temperature $T_b$ and the restitution coefficient $\alpha$.  To give a
flavour of the regime we are working on, in Fig.~\ref{fig1} we
represent two typical configurations of particle positions in the
system from simulations with different restitution coefficients.  To
fix ideas, useful for the following discussion, we now focus on a
particular choice of parameters. The system has height
$L_y/\sigma=600$ and width $L_x/\sigma=180$, the number of particles
is $N=1500$ (corresponding to a number of resting layers
$N_{r}=N\sigma/L_x\approx 8.3$), and the restitution coefficient is
$\alpha=0.95$. Gravity acceleration is set to $g=9.8$,
while $T_b/(gL_y)=2.55$.

\begin{figure}[htbp]
\begin{center}
\includegraphics[angle=0,width=8.5cm,clip=true]{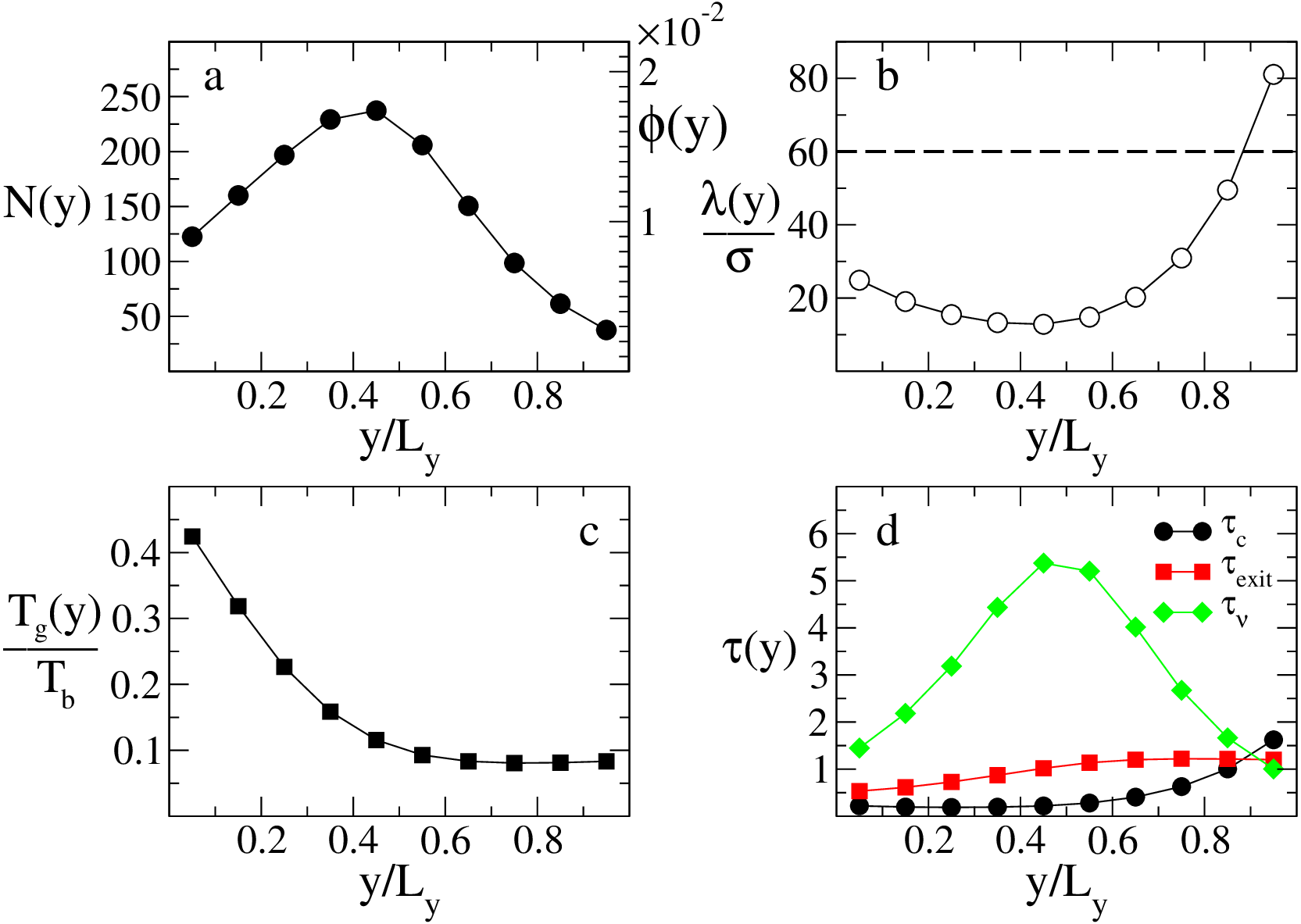}
\caption{The main local observables as function of the rescaled
  distance $y/L_y$ from the bottom wall: a) the local number $N(y)$ of
  particles and the local packing fraction $\phi(y)$ (see right $y$ scale);
 b) the local mean free path $\lambda(y)$; c)
  the local rescaled granular temperature $T_{g}(y)/T_b$; d) local
  characteristic times (see text for details). The other parameters
  are $N=1500$, $\alpha=0.95$ and $T_b/(mgL_y)=2.55$.
\label{fig2}}
\end{center}
\end{figure}

In Figure~\ref{fig2} the behavior of the main ``local'' observables is
reported versus the distance from the floor. We have divided the total
height into $m_l=10$ horizontal layers of height $L_l/\sigma=60$, so
that $m_lL_l=L_y$. In each layer (whose height $y$ is measured at its middle point) we have computed the average of some
observable which is expected to be {\em slowly} varying, i.e. the
number of particles in the layer  $N(y)$ and the
average kinetic energy in that layer $T_{g}(y)\equiv
\frac{1}{2}[T_{x}(y)+T_{y}(y)]$, which are directly associated to the
density and temperature hydrodynamic fields ~\cite{BDKS98}.
In Figures~\ref{fig2}a and~\ref{fig2}c we show $N(y)$ and $T_{g}(y)$
respectively. In Fig.~\ref{fig2}a one can also appreciate (see right
scale) the local packing fraction $\phi(y)=n(y)\pi\sigma^2/4$
where $n(y)=N(y)/(L_xL_l)$ is the local density.  We also give further
information to assess the granular regime chosen for our study. In
Figure~\ref{fig2}b we report the behavior of the local mean free path
$\lambda(y)=(n(y)\sigma)^{-1}$ (note that $\lambda(y)$ is proportional
to the local mean free path through a order $1$ geometrical factor and
through the Enskog constant $g_2(\sigma)$ which, in our dilute case,
is very close to $1$). Finally in Fig.~\ref{fig2}d some local
characteristic times are reported for the different layers:
$\tau_{c}(y)=\lambda(y)/v_{th}(y)$ is the mean collision time,
$\tau_{exit}(y)=L_l/v_{th}(y)$ is the mean exit time from the layer
with $v_{th}(y)$ equal to the local thermal velocity and
$\tau_{\nu}(y)$ is the time associated to vorticity diffusion obtained
from the expectation for the local kinematic viscosity $\nu(y)$, taken
from the granular Enskog theory~\cite{BDKS98}. These times will be
useful (and will be explained in detail) in the discussion of Section
IV.


We have verified that the choice of the system parameters as well as
the choice of $m_l$ meet the following constraints:
\begin{itemize}
\item {\em Stability of horizontal modes}: if the system is too large
  it is known that instabilities can arise, starting with shear modes
  and then involving clustering~\cite{GZ93,NEBO97}; a minimum
  criterion to avoid that is enforcing $k_{min}> k_{\perp}(y) \equiv
  \sqrt{\zeta_{H}(y)/(2\nu(y))}$ with $\zeta_{H}(y)$ the zero-th order
  approximation of the homogeneous cooling rate in a layer,
  and $\nu(y)$ the kinematic viscosity in the layer; because
  $k_{\perp}(y) \propto 1/\lambda(y)$ this condition determines a
  constraint on the maximum number of the particles in each layer.
\item {\em Absence of convection rolls}: it is known that this
  granular setup is also subject to thermal convection instability due
  to the competition between gravity and temperature
  gradients~\cite{RRC00,WHP01}; by direct inspection of the velocity
  field and the dynamical evolution of the numerical simulation we
  have verified that no convection is present in our system.
\item {\em Fast local relaxation of microscopic modes}: this condition
  is equivalent to ask that a particle in a layer suffers many
  collisions before going out from the layer, which (in view of the
  fact that $L_l<L_x$) is equivalent to $\lambda(y)<L_l$ (note that
  this is not strictly verified in the topmost layer).
\item {\em Diluteness}: to avoid strong velocity correlations and hope
  a reasonable comparison with dilute hydrodynamic granular theories,
  we also require that the local packing fraction is small,
  i.e. $\phi(y) \ll 1$.
\item {\em Local homogeneity}: finally, a major constraint in the
  choice of $m_l$ is given by ``quasi-homogeneity'' inside the $i$-th layer,
  i.e. we require that $[a_{i+1}-a_i]^{-1}a_i \ll 1$, being $a_i$ one
  of the ``slow'' variables, i.e. $N$ or $T_{g}$.
\end{itemize}
After having checked by direct inspection that all these requirements
are satisfied in our system and with this choice of $m_l$, we have to
warn the reader about two important properties which distinguish this
one with respect to other setups, where fluctuating hydrodynamics has
been previously discussed (e.g. the homogeneous cooling
state~\cite{BGM08}). Each horizontal layer is considered here as a
single finite system where the fluctuations of velocity shear mode is
studied. Anyway:
\begin{enumerate}
\item
Each layer is an {\em open} system, exchanging particles with adjacent
layers; we will discuss how this peculiarity do affect the studied
fluctuations, in particular the autocorrelation decay.
\item
Tuning the external parameters, such as $\alpha$, the total number of particles $N$ or the
parameters of the thermal base $T_b$, affects the properties of each
layer, e.g. local density or temperature, in a complicate and not
direct way. Even if solutions of the hydrodynamic equations for this
particular $2D$ setup exist~\cite{BRM01}, the boundary layer near the
thermal base always requires a careful treatment and makes
quantitative predictions for local variables, from the only knowledge
of external parameters, quite hard.
\end{enumerate}

\section{Numerical results for hydrodynamic fluctuations}
\subsection{Time decay of the autocorrelation and dependence on wave number}
\begin{figure}[htbp]
\begin{center}
\includegraphics[angle=0,width=7.cm,clip=true]{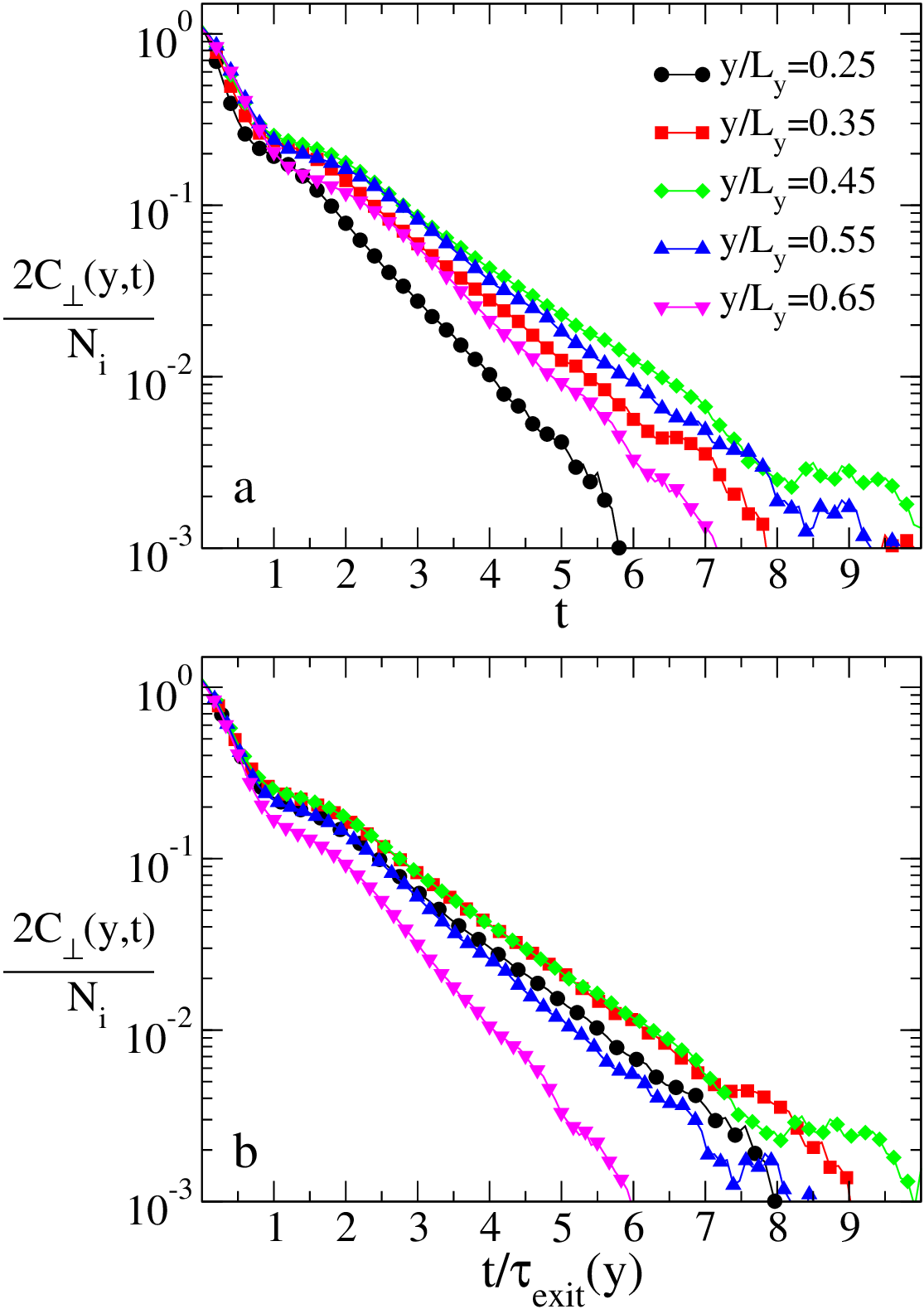}
\caption{The local rescaled correlation function $C_{\perp}(y,t)$ as
  function of the time $t$ (a) and rescaled time $t/\tau_{exit}(y)$ (b)
  for different layers. The other parameters are the same of
  Fig. \ref{fig2}.
\label{fig3}}
\end{center}
\end{figure}

The behavior of $C_{\perp}(y,t)$ obtained in layers at different height $y$, from the
simulations, is shown in Fig.~\ref{fig3}a. The autocorrelation function
of the shear mode presents two main regimes: a first rapid decay
toward an intermediate plateau and a second decay to zero, which is
roughly exponential.

Our interpretation of the first decay is that it is associated to the
openness of the layer: particles belonging to the layer at time $t=0$
escape from it with a typical velocity $\vt(y)=\sqrt{2T_{g}(y)}$, being
on average replaced by particles coming from the adjacent layers. In
Fig.~\ref{fig3}b we have rescaled the time with
$\tau_{exit}(y)=L_l/\vt(y)$, getting a fair collapse of the first decay
in different layers. This collapse confirms that the typical time of
this first decay is $\sim \tau_{exit}(y)$ and that this is associated
to the exchange of particles through the boundaries of the layer. The
second decay can be reasonably fit by an exponential law. This fit
works better in the central layers,
i.e. $0.25<\frac{y}{L_y}<0.65$. Problems in the top and bottom
regions could be related to the mean free path becoming too large
(this is particularly true in the top region, where it gets closer to
the width of the box): in that cases the mean free time could become
close to the time taken by a sound wave to travel along the width of
the system, producing non-exponential relaxation which could also be
size-dependent. The region near the base is also affected by the
complicate boundary layer where particles take an asymmetrically
distributed vertical velocity because of the thermostat. From now on
we focus on the central layers where the decay is well fit by $C_\perp
\sim \exp(-t/\tau(y))$.
\begin{figure}[htbp]
\begin{center}
\includegraphics[angle=0,width=6.5cm,clip=true]{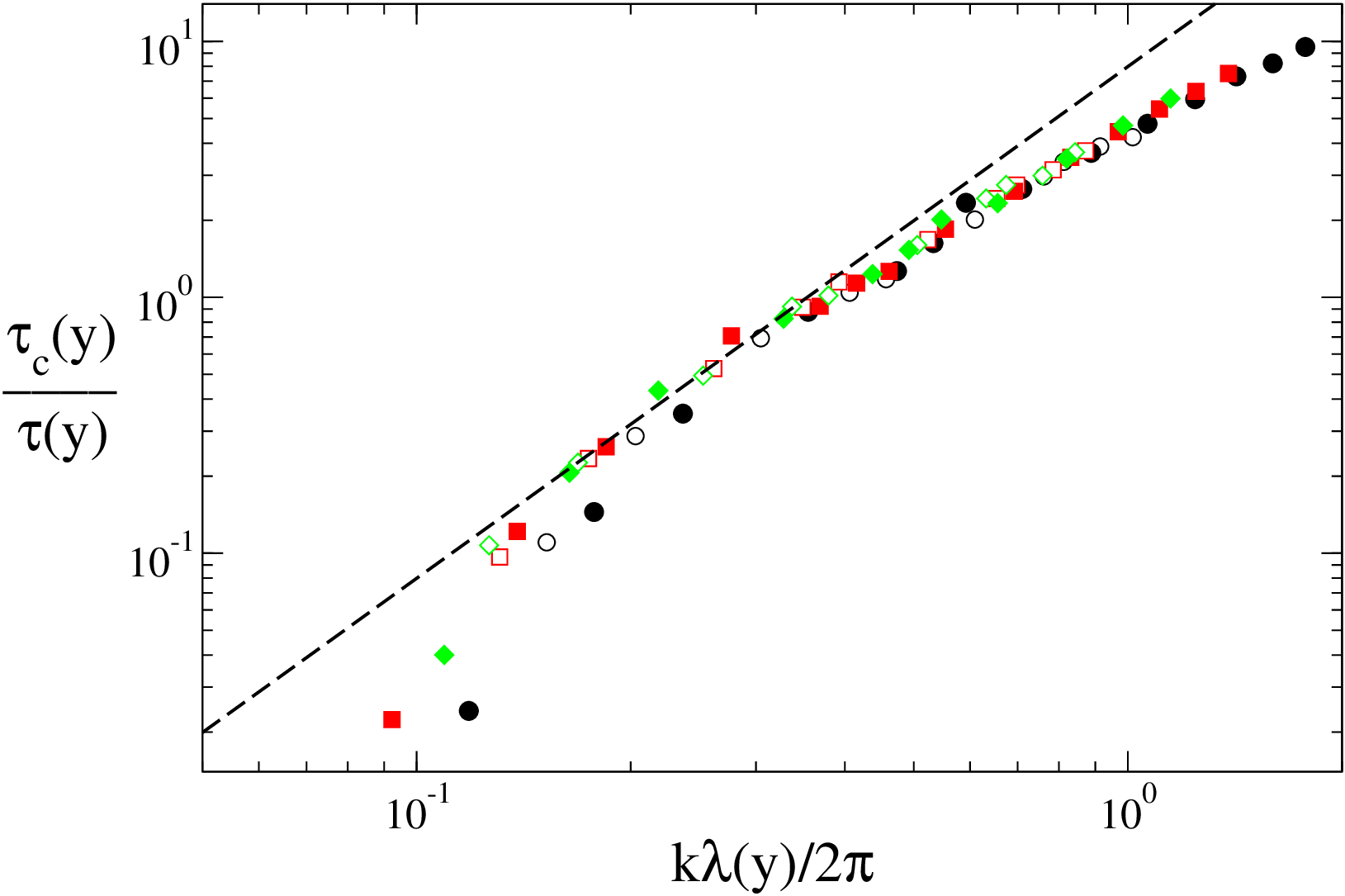}
\caption{ The inverse of the decay time $\tau(y)$ as a function of the wave number $k$. The ordinate is been rescaled by the mean collision time $\tau_{c}(y)=\lambda(y)/v_{th}(y)$, while the wave number is rescaled by the local mean free path $\lambda(y)/2\pi$. The data have a good collapse for the two analyzed total densities $n\sigma^2=1.4\cdot 10^{-2}$ (open symbols) and $n\sigma^2=1.7\cdot 10^{-2}$ (closed symbols) and for three different layers: a) $y/L_y=0.25$ (circles), b) $y/L_y=0.35$ (squares) and c) $y/L_y=0.45$ (diamonds). For completeness
the theoretical prediction (dashed curve) of the local decay time $\tau_{\nu}(y)$, due to the vorticity diffusion, is shown (see text for details). The other parameters are the same of Fig. \ref{fig2}.
\label{fig4}}
\end{center}
\end{figure}

In those cases, it is interesting to study the decay time $\tau(y)$ as
a function of $k$ and to compare it with the theoretical prediction
associated to vorticity diffusion $\tau_{\nu}(k,y)=1/(\nu(y) k^2)$
which describes the decay of shear modes autocorrelation in a
homogeneous elastic system with the viscosity of the layer $\nu(y)$
(see Eq.~\eqref{eqnu}). To this aim we have performed some simulations
modifying only the horizontal size $L_x$ of the box and maintaining
the total density $n$ constant. In this way we can obtain a better
resolution in $k$. Our analysis is focalized to the total density
$n\sigma^2=1.4\cdot 10^{-2}$ (the same of Fig. \ref{fig2} and
Fig. \ref{fig3}) and $n\sigma^2=1.7\cdot 10^{-2}$ and the comparison
is shown in Fig.~\ref{fig4}. Our first observation is that all data,
appropriately rescaled, collapse indicating the same dependence on the
local parameters $n(y)$ and $T_{g}(y)$ as the homogeneous system. In
particular, data in the range $0.1<k\lambda(y)/(2\pi)<0.5$ show a very
good agreement with the prediction of simple vorticity diffusion due
to shear viscosity $\tau(y) = \tau_{\nu}(k,y)=1/(\nu(y) k^2)$. From
values $k\lambda(y)/(2\pi)\approx 0.7$, $\tau(y)$ shows deviations from
$\tau_{\nu}(y)$: we suspect that this is an indication of the failure
of the hydrodynamical approach at such small lengthscales. A less
clear failure is present also at very large wavelengths
($k\lambda(y)/(2\pi)<0.1$): such a discrepancy suggests the presence of
some mechanism slowing the relaxation of shear modes, perhaps a
precursor of power-law relaxation associated to the two-dimensional
geometry.
 

\subsection{Structure factor of transverse velocity modes}

The amplitude of fluctuations of the shear mode $U_{\perp}(k,y,t)$ in
a layer, which is directly proportional to the transverse
velocity structure factor, is given (see Eq.(\ref{Uk})) by
\begin{equation} 
C_{\perp}(k,y,0) = \frac{\langle U_{\perp}(0)U^*_{\perp}(0)\rangle}{\vt^2(y)}=
\frac{1}{\vt^2} \Big[\langle \sum_{j=1}^{N(y)} v_{y,j}^2\rangle+ \langle\sum_{j=1}^{N(y)} \sum_{\substack{p=1 \\ p\neq j}}^{N(y)} v_{y,j} v_{y,p} e^{-i k \pi (X_j-X_p)}\rangle\Big]=\frac{N(y)}{2}\Big[1+\Delta(k,y)\Big].
\label{corr}
\end{equation}
From Eq.~\eqref{corr} it is seen that $C_{\perp}(k,y,0)$ can be split
into two parts: a first {\em self} term which scales as $\sim N(y)$ and
a second term, $\sim N(y)\Delta(k,y)$ that contains the informations of
the $k$-dependence in a layer at height $y$.

\begin{figure}[htbp]
\begin{center}
\includegraphics[angle=0,width=7cm,clip=true]{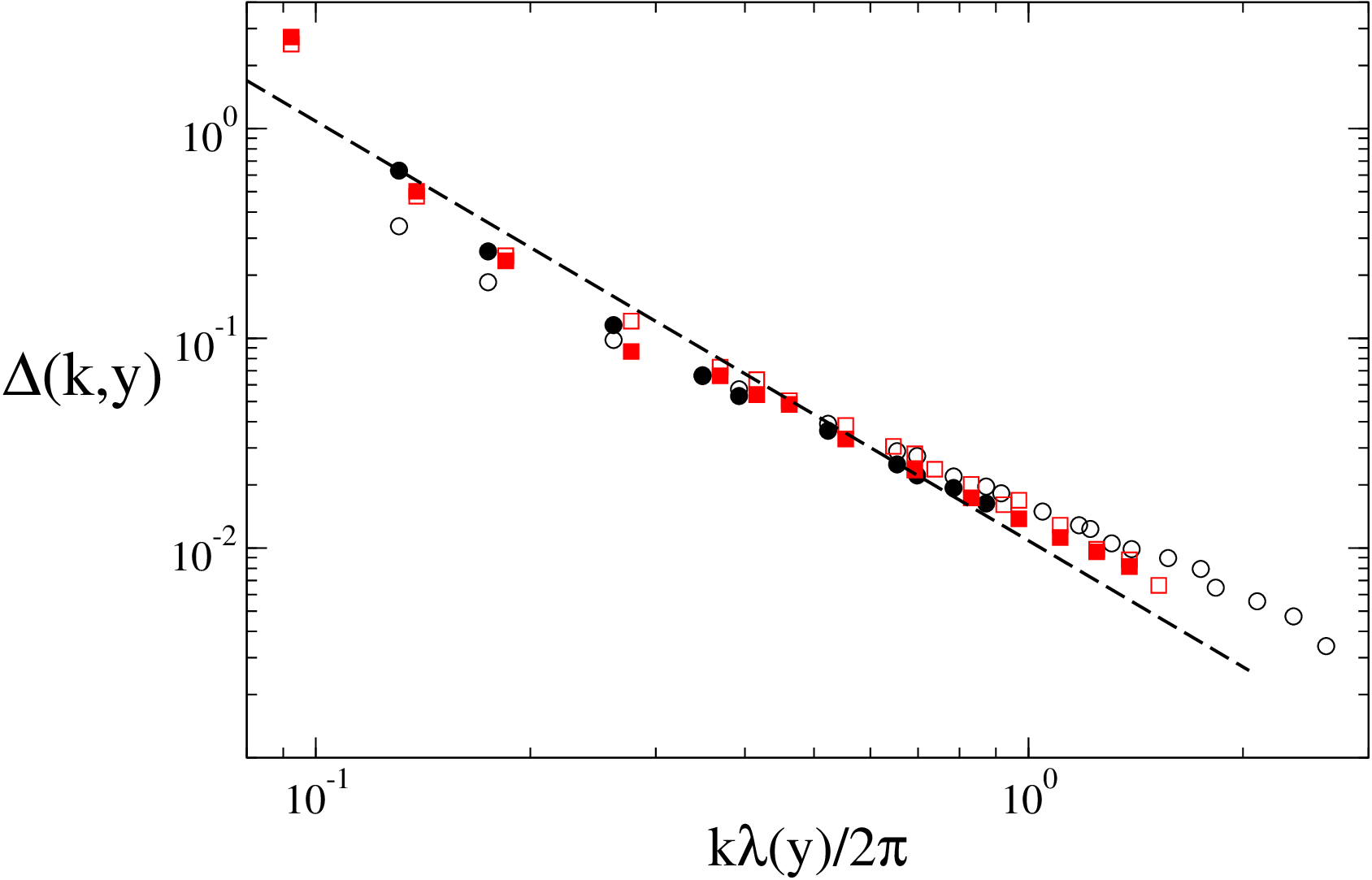}
\caption{ The local quantity $\Delta(k,y)$ (see Eq. (\ref{corr}) for a definition) as a function of the wave number $k$ rescaled by the local mean free path $\lambda(y)/(2\pi)$. The open symbols are the simulation data for the fixed layer with $y/L_y=0.35$ and for two different total densities a) $n\sigma^2=1.4\cdot 10^{-2}$ (open circles) and b) $n\sigma^2=1.7\cdot 10^{-2}$ (open squares). The closed symbols are the corresponding values of the two densities a) and b), obtained from Eq. (\ref{eqCk}) using an exponential fit for the decay times $\tau(y)$ (data plotted in Fig. \ref{fig4}). The collapsing data are compared to the theoretical prediction (dashed curve) using in Eq. (\ref{eqCk}) the local decay time $\tau_{\nu}(y)$ due to the vorticity diffusion. The other parameters are the same of Fig. \ref{fig2}.
\label{fig5}}
\end{center}
\end{figure}
In a homogeneous equilibrium fluid, from equilibrium statistical
mechanics, one expects the absence of spatial velocity correlations,
corresponding to a flat $C_\perp(k,y,0)$, as expressed by
Eq.~\eqref{corrLL}. Homogeneously cooling granular systems (with size
smaller than the critical size for linear stability of shear modes)
are expected to present a non-flat structure factor, implying spatial
velocity correlations, as marked by Eq.~\eqref{corrHcs}.  However
those systems present an instability which is not directly observed in
our simulations (unless one gets much larger horizontal size
$L_x$). Moreover, our layer is in a stationary state, a regime quite
different from the cooling state: such a difference is reasonable
because each layer is ``driven'' by particles coming from nearby
layers. The best candidate as a model for quantitative comparison with
our numerical results is the homogeneously driven system studied
in~\cite{NETP99}, whose fluctuating hydrodynamics we have resumed in
Eq.~\eqref{LangevinTrizac}.  Our system is different from that model,
since we have not a uniform noise throughout the system, but the
addition of energy is localized at the bottom of the
box. Nevertheless in a single layer there is a balance between the
energy exchanged with adjacent layers and the energy dissipated in
collisions. It is therefore tempting to introduce an effective
external noise whose amplitude is determined by the energy balance and
is proportional to $N(y)\zeta_{H}(y) T_{g}(y)$. Using this expression
for noise and considering that the characteristic decay time is
$\tau(k,y)$, the quantity $C_{\perp}(k,y,0)$ can be written as
\begin{equation}
C_{\perp}(k,y,0)=\frac{N(y)}{2}\Big[ 1+\frac{\zeta_{H}(y)}{2}\tau(k,y)\Big].
\label{eqCk}
\end{equation}
Comparing Eqs. (\ref{corr}) and (\ref{eqCk}) we note that
$\Delta(k,y)$ can be written as $\zeta_{H}(y)\tau(k,y)/2$. In an
elastic homogeneous system this term is zero, while in a granular
system we observe a non-negligible contribute of $\Delta(k,y)$.  In
order to verify our assumption we have plotted in Fig. \ref{fig5} the
simulation data for $\Delta(k,y)$ and its prediction based on the
Eq. (\ref{eqCk}) and estimated using for $\tau(y)$ the data plotted in
Fig. \ref{fig4}.  The agreement is very good for both the cases
analyzed (open and closed symbols in Fig. \ref{fig5}) and this
validates our considerations. Moreover we have verified that the
probability distribution of $U_{\perp}(k,y,t)$ is close to a Gaussian, consistent with a linear Langevin
equation.  In Fig. \ref{fig5} only results from the central layer
($y/L_y=0.35$) are shown: however we have verified that it is equally
good in the contiguous layers. For the sake of completeness we have
shown in the figure also $\Delta(k,y)$ obtained using as
characteristic time $\tau_{\nu}(y)$. The disagreement is small and analogous to
that already seen in the previous section.


\subsection{Amplitude of fluctuations for the mode at largest wavelength}

To have a better assessment of our hypothesis (Eq.~\eqref{eqCk}), we
focus now on the amplitude $C_{\perp}(k_{min},y,0)$ of the shear mode
for the largest available wavelength $k_{min}$. Indeed one expects
that at larger wavelengths hydrodynamics works better. For such a
purpose, we have fixed the width $L_x$ ($L_x/\sigma=180$) and we have
changed the total number $N$ of the particles.  In this way the local
density $n(y)$ is also changed and we can explore the dependence of
$\Delta(y)\equiv \Delta(k_{min},y)$ on the particle number or,
equally, on $1/\lambda(y)$.\\
\begin{figure}[htbp]
\begin{center}
\includegraphics[angle=0,width=7.5cm,clip=true]{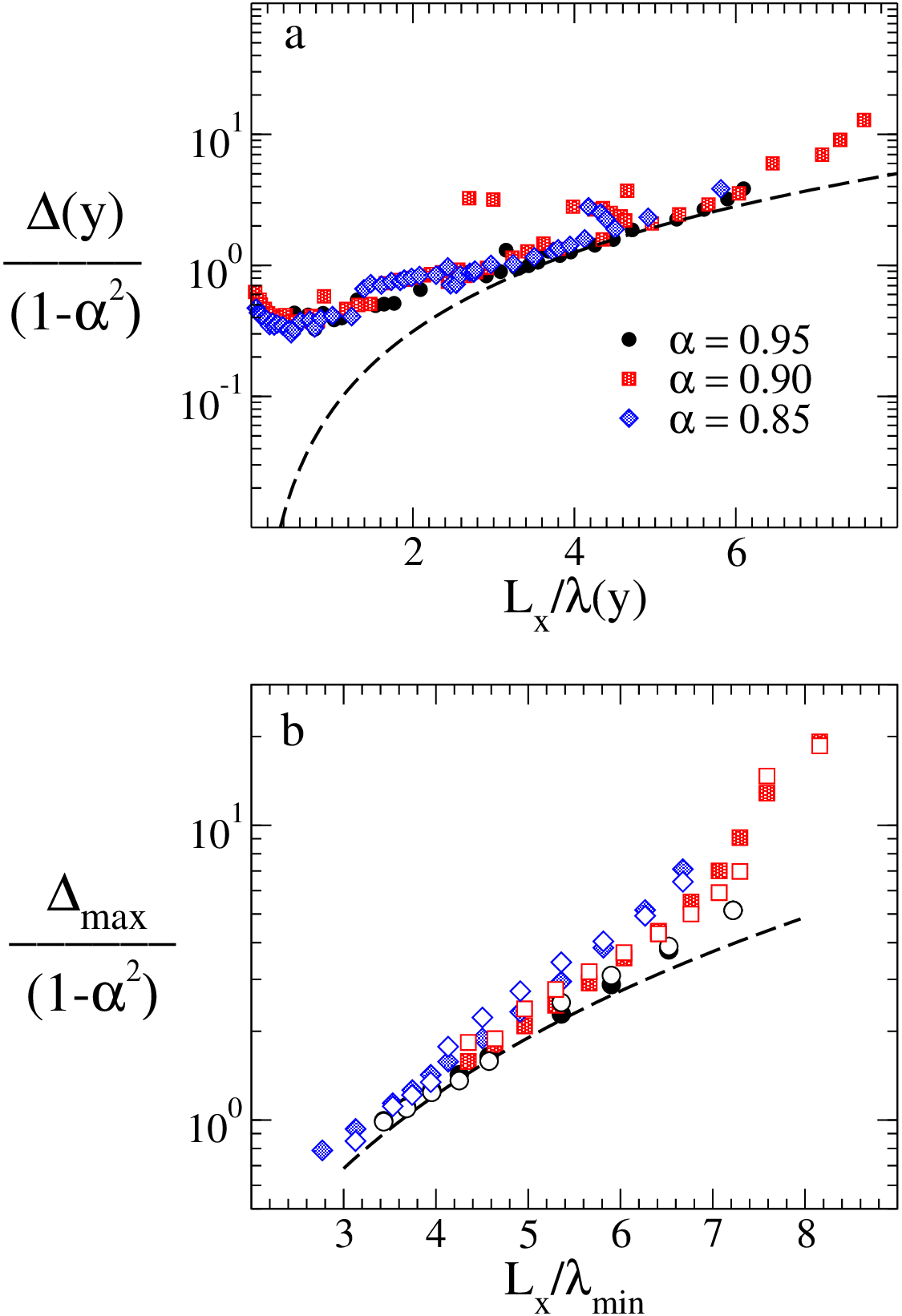}
\caption{ Panel a: The parameter $\Delta(y)\equiv
  \Delta(k_{min},y)$ rescaled by $1-\alpha^2$ as function of the
  inverse of the rescaled local mean free path $L_x/\lambda(y)$ for
  different inelasticities: $\alpha=0.95$ (circles), $\alpha=0.90$
  (squares) and $\alpha=0.85$ (diamonds). The dashed curve is the
  quantity $\zeta_{H}(y)/(2\nu(y) k^2_{min})$ corresponding to the
  homogeneous cooling theory (see
  Eqs. (\ref{LangevinHcs})-(\ref{eqzeta})) Panel b: The parameter
  $\Delta_{max}$ (see text for a definition), rescaled by
  $1-\alpha^2$, as function of the inverse of the $L_x/\lambda_{min}$,
  with $\lambda_{min}=(n^{max}(y)\sigma)^{-1}$, for the same
  inelasticities of panel a (closed symbols). The open symbols
  correspond to the data obtained from Eq. (\ref{eqCk}) (see text for
  details). The dashed curve is the same of panel a valued on the
  layer with the maximum of the density.  The other parameters of the
  two panels are the same of Fig.\ref{fig2}.
\label{fig6}}
\end{center}
\end{figure}
Measures of $\Delta(y)$ for many different layers and for systems with
three values of the restitution coefficient $\alpha$ ($0.85$, $0.90$
and $0.95$) are shown in in Fig.~\ref{fig6}a. The data are rescaled by
the factor $1-\alpha^2$ because this term contains the main dependence
on $\alpha$ in the Eq. (\ref{eqCk}). The collapse of data is quite
remarkable, if one consider that data come from different systems and
from layers which can be at very different distance from the base: for
instance, a given value of $n(y)$ can correspond both to layers below
the density maximum and above it. At moderate values of $n(y)$ the
data are closed to the values obtained using $\tau_{\nu}(y)$ as decay
time in Eq. (\ref{eqCk}). The disagreement at small values of $n(y)$
is due to the fact that hydrodynamic hypothesis is not longer valid in
this range.  It is likely that the deviations from a clean collapse in
Fig.~\ref{fig6}a are due to the mixing of data from regions
characterized by too different physical parameters as $T_g$ and
$\alpha$. We have repeated the analysis by focusing only on the layer
containing the maximum of the density $n^{max}(y)$, but changing the
external parameters in order to explore different values of
$n^{max}(y)$. The corresponding value $\Delta_{max}$ is presented in
Fig.~\ref{fig6}b. The data do not rescale exactly and this indicates
an uncleared $\alpha$ dependence. On the other side the measures are
in good agreement with the corresponding data obtained from
Eq. (\ref{eqCk}) (open symbols in figure). This suggests that the
Eq. (\ref{eqCk} ) reproduces the right behavior provided that the
suitable decay time is used. Concerning this point, it is important to
notice that for large values of $n(y)$, $\Delta(y)$ (or
$\Delta_{max}$) can become much larger than that observed in the
homogeneous cooling state (see dashed line in Fig. \ref{fig6}).

\section{Conclusions}

We have analyzed the fluctuations of the shear mode of the velocity in
nearly homogeneous subregions of a inhomogeneous granular fluid. Our
conclusion is that - neglecting regions too close to the upper and
lower boundaries - the large time temporal decay of their
autocorrelation does not deviate dramatically from the expectation for
a homogeneous equilibrium fluid, i.e. $1/\nu(y) k^2$, provided that
hydrodynamic (large enough) scales are considered. At too large
scales, comparable with the total horizontal size, some deviations
from the $1/\nu(y) k^2$ behavior is appreciated, likely due to finite
size effects and problems associated to the 2D geometry. \\ The study
of the amplitude of fluctuations, i.e. the transverse velocity
structure factor, suggests that for the central layers, the effective
noise is well described by a sum of two contributions, both white and
Gaussian: an internal one, associated to the vorticity current, which
is at ``equilibrium'' with the local temperature, and an external one
which is responsible for the balance of local energy which is
continuously lost in collisions. We have shown that those assumptions
lead to a good estimate of the structure factor in all cases analyzed,
if the autocorrelation decay time used in the formula is the one
measured in the simulations: such an agreement holds even when the
decay time deviates from the simple prediction $1/\nu(y) k^2$.\\ We
stress that formula~\eqref{eqCk} for the amplitude of fluctuations
represents an explicit violation of the equilibrium
Fluctuation-Dissipation Relation, which would yield a constant
structure factor: such a violation can be quite large with respect to
those discussed in the homogeneous cooling regime~\cite{BGM08}. In
particular, violations observed here lead to an amplitude of
fluctuations which can be {\em much larger} than that expected at
equilibrium or in the homogeneous cooling theory. It is also
remarkable to notice that the correlations observed here are not
related to the gradients in the system (as it happens for other
correlations in molecular systems with an imposed thermal or density
gradient), but are a direct consequence of inelastic collisions.

\section*{Acknowledgments} We acknowledge Prof. U. Marini Bettolo Marconi and Dr. G. Gradenigo for useful discussions and a critical reading of the manuscript. The work of the authors is supported by the ``Granular-Chaos'' project, funded by the Italian MIUR under the FIRB-IDEAS grant number RBID08Z9JE. 


\bibliography{fluct.bib}


\end{document}